\documentclass[pdflatex,sn-mathphys-num]{sn-jnl}

\usepackage{graphicx}%
\usepackage{multirow}%
\usepackage{amsmath,amssymb,amsfonts}%
\usepackage{amsthm}%
\usepackage{mathrsfs}%
\usepackage[title]{appendix}%
\usepackage{xcolor}%
\usepackage{textcomp}%
\usepackage{manyfoot}%
\usepackage{booktabs}%
\usepackage{algorithm}%
\usepackage{algorithmicx}%
\usepackage{algpseudocode}%
\usepackage{listings}%

\theoremstyle{thmstyleone}%
\theoremstyle{thmstyletwo}%

\theoremstyle{thmstylethree}%

\usepackage{booktabs}
\usepackage{graphicx}
\usepackage{array}
\usepackage{placeins}
\usepackage{tikz}
\usetikzlibrary{arrows.meta,positioning}
\usepackage{amsmath,amssymb}
\usepackage{xcolor}
\definecolor{figureblue}{RGB}{23,54,93}
\definecolor{figurefill}{RGB}{243,246,249}

\usepackage{pifont}
\usepackage{xspace}

\usepackage{todonotes}

\definecolor{lightred}{RGB}{255, 200, 200}  
\definecolor{lightblue}{RGB}{200, 220, 255} 
\definecolor{mycolor}{rgb}{0.902,0.902,0.980}
\definecolor{darkgreen}{rgb}{0.2,0.8,0.25}

\usepackage{float}

\newcommand{\standardtablestyle}{%
    \small
    \renewcommand{\arraystretch}{1.12}
    \setlength{\tabcolsep}{4pt}
}

\begin{document}

\title[From Prediction to Action: An Execution-Aware End-to-End Machine Learning Framework for Structured Investment Decision Support]{From Prediction to Action: An Execution-Aware End-to-End Machine Learning Framework for Structured Investment Decision Support}


\author[1]{Aditya Bhardwaj} \email{ady.bhardwaj2004@gmail.com} 

\author*[1]{Xingchen Zhang}\email{x.zhang12@exeter.ac.uk}

\affil*[1]{Department of Computer Science, University of Exeter, Exeter, EX4 4RN, UK}


\abstract{Financial machine learning is widely used for market prediction, but many existing studies formulate trading as a directional forecasting problem and stop at predicting whether prices will rise or fall. Such predictions are difficult to translate directly into practical trading decisions, which also require execution rules, risk controls, and clearly defined actions. To address this gap, we propose an end-to-end machine learning framework for structured investment decision support. The framework integrates multi-asset market data, execution-aware label construction, feature engineering and refinement, specialist predictive models, and ensemble learning. Rather than defining learning targets solely according to future price direction, the proposed execution-aware formulation considers realised trading outcomes under predefined stop-loss and take-profit rules. Model outputs are subsequently converted into structured trading recommendations comprising trade direction, entry price, stop-loss, take-profit, and risk–reward ratio. The framework is evaluated using time-aware forward testing, with results showing that execution-aware labelling improves risk-adjusted trading performance over conventional directional labelling and that the integrated framework outperforms a simpler single-model baseline across key trading metrics. An automated cloud-based implementation further demonstrates the operational feasibility of the proposed approach. Overall, the findings suggest that closer alignment between predictive modelling, execution logic, and structured decision generation can improve the practical value of machine learning for financial trading.}

\keywords{Financial machine learning \sep algorithmic trading \sep execution-aware labelling \sep ensemble learning \sep decision support \sep cloud deployment}

\maketitle

\section{Introduction}

Financial markets generate enormous volumes of sequential and heterogeneous data, creating substantial opportunities for machine learning to support investment decision-making. Over the past decade, machine learning has been applied to stock-price prediction, foreign-exchange forecasting, empirical asset pricing, portfolio management, and algorithmic trading \citep{patel2015,bao2017,gu2020,shen2021,masini2023}. Recent reviews document rapid growth in deep learning, hybrid modelling, and artificial-intelligence-based trading across multiple financial markets \citep{chen2024review,dakalbab2024}. Recent empirical evidence also demonstrates the potential of nonlinear models. For example, \citet{htun2024} evaluated random forests, support vector machines (SVM), and long short-term memory (LSTM) networks across 494 S\&P 500 stocks and found that the LSTM model achieved the strongest classification performance. Collectively, these developments establish machine learning as an important research direction in quantitative finance, but they do not fully resolve how predictive improvements should be converted into dependable trading decisions.

First, practical trading differs fundamentally from an isolated forecasting exercise. Many studies continue to formulate financial prediction as directional classification or regression, where the objective is to estimate whether an asset price will increase or decrease over a future horizon \citep{wu2020,chen2024review}. Recent empirical work also evaluates whether returns will exceed a predefined threshold \citep{htun2024}. Although these formulations support convenient model comparison, they capture only part of the decision problem. Research on reinforcement learning in finance shows that practical financial decision-making also involves optimal execution, portfolio allocation, market making, hedging, and risk-sensitive control \citep{hambly2023}. A useful trading system must therefore determine not only the expected price direction but also the entry price, acceptable risk, stop-loss and take-profit levels, and whether the potential reward justifies the exposure.

Second, evaluation methodology creates another important difficulty. Many studies emphasise prediction-oriented measures such as accuracy, precision, recall, or F1 score, whereas practical trading performance depends on risk-adjusted outcomes such as Sharpe ratio, maximum drawdown, profit factor, and trade expectancy \citep{lopez2018,hwang2023,fu2024}. Recent benchmarking research additionally identifies low signal-to-noise ratios, survivorship bias, and backtesting overfitting as central challenges in financial machine learning \citep{liu2022finrlmeta}. Moreover, training labels are frequently defined according to final price direction rather than realised outcomes under predefined execution rules. A mismatch can therefore arise between the objective optimised during model training and the criteria used to determine whether a trade would have been successful in practice.

Additionally, operational implementation at the system level also remains comparatively underdeveloped. A recent systematic review of 143 studies found that only 16\% of the reviewed financial-trading applications entirely automated the trading process \citep{dakalbab2024}. A further recent review highlights continuing limitations involving usability, standardisation, domain-knowledge integration, and the practical implementation of advanced financial forecasting models \citep{chen2024review}. Initiatives such as FinRL-Meta demonstrate the value of automated data curation, reproducible market environments, benchmarking, and cloud deployment \citep{liu2022finrlmeta}. Nevertheless, the wider literature still gives limited attention to connecting data preparation, execution-aware prediction, risk-defined trade construction, human-readable reporting, and scheduled delivery within one reproducible decision-support workflow.

Motivated by these limitations, we propose an end-to-end machine learning framework for structured investment decision support. An execution-aware formulation aligns the learning objective with realised trading outcomes under predefined execution rules, rather than treating trading as a simple directional prediction problem. A unified pipeline brings together multi-asset market data, execution-aware label construction, large-scale feature engineering, feature refinement, specialist predictive models, ensemble learning, and automated report generation. Its outputs extend beyond directional predictions to include trade direction, entry price, stop-loss, take-profit, and risk--reward ratio. Cloud-based execution and scheduled daily prediction further support repeated operational use. Additionally, this 
study does not automate order execution; instead, it addresses operational integration while retaining human review of the generated recommendations. 

The main contributions of this work are summarised as follows:
\begin{itemize}

\item We introduce an execution-aware learning formulation that aligns model training with realised trading outcomes under predefined execution rules, reducing the mismatch between directional prediction and practical trading decisions.

\item We develop an end-to-end machine learning framework that integrates multi-asset market information, feature refinement, specialist predictive models, and ensemble learning to generate structured investment recommendations.

\item We systematically evaluate the proposed framework using time-aware forward testing and trading-oriented performance measures, including comparisons with conventional directional labelling and a simpler single-model baseline.

\item We demonstrate the operational feasibility of the framework through automated decision-support generation and cloud-based deployment while retaining human review of the generated recommendations.

\end{itemize}

The remainder of this paper is organised as follows. Section 2 reviews related work in financial machine learning, execution-aware prediction, and operational decision-support systems. Section 3 presents the proposed framework in detail. Section 4 describes the evaluation protocol, reports the experimental results, and discusses the main findings. Section 5 outlines the limitations of the study and directions for future research. Finally, Section 6 concludes the paper.

\section{Related Work}
\subsection{Machine Learning for Financial Prediction}

Machine learning has become an important research direction in financial prediction because financial markets generate complex, noisy, and highly non-stationary time series that are difficult to model using traditional statistical approaches. Compared with linear forecasting models, machine learning methods are capable of capturing nonlinear relationships, exploiting high-dimensional feature spaces, and integrating heterogeneous sources of market information. Consequently, these methods have been widely applied to stock-price prediction, exchange-rate forecasting, asset pricing, and portfolio management. Numerous studies have reported encouraging predictive performance using both conventional machine learning algorithms and deep learning models \citep{patel2015,bao2017,shen2021,gu2020}.

Despite these advances, the reported performance improvements should be interpreted with caution. Financial prediction results are often highly sensitive to the choice of dataset, input features, evaluation period, and benchmark model, making it difficult to determine whether the observed gains reflect genuine predictive capability or favourable experimental settings. As discussed by \citet{masini2023}, although machine learning has substantially advanced time-series forecasting, its superiority often depends on the characteristics of the prediction task and does not apply universally across financial markets. Accordingly, improvements reported in the literature should be evaluated not only by predictive performance but also by their robustness under changing market conditions.

Another important direction in the literature is the development of ensemble and hybrid learning strategies. By combining multiple models or complementary sources of information, ensemble methods aim to improve predictive robustness and reduce the dependence on any individual learning algorithm \citep{rapach2010,jiang2020}. Similarly, recent deep learning approaches have demonstrated strong representation-learning capabilities for financial forecasting tasks \citep{bao2017}. However, these increasingly sophisticated models also introduce new challenges. Their performance often depends heavily on feature engineering, hyperparameter selection, and the specific market regime used for evaluation, while their limited interpretability makes it more difficult to understand the underlying decision process. Consequently, improvements in predictive accuracy do not necessarily translate into more reliable or practically useful trading decisions.

Overall, the existing literature demonstrates that machine learning provides a powerful set of tools for financial prediction and can outperform traditional forecasting approaches under appropriate conditions. Nevertheless, most existing studies remain primarily prediction-oriented, with research focusing on improving forecasting accuracy or return prediction rather than supporting practical investment decision-making. As a result, relatively little attention has been paid to how predictive models can be integrated into complete decision-support systems that operate under realistic trading constraints. The resulting gap motivates the execution-aware perspective discussed in the following subsection.

\subsection{Execution-Aware Learning for Trading}

A fundamental limitation of many financial machine learning studies lies in the definition of the prediction target. Most existing methods formulate trading as a directional classification or regression problem, where the objective is simply to predict whether the future price will increase or decrease over a predefined horizon \citep{wu2020}. Although this formulation is convenient for model development and comparison, it does not accurately reflect how trading decisions are evaluated in practice. A successful trade depends not only on the eventual price direction, but also on whether predefined stop-loss or take-profit levels are reached first, how the price evolves throughout the holding period, and whether the trade satisfies acceptable risk--reward requirements. Consequently, directional labels often introduce a mismatch between the optimisation objective during model training and the criteria used to evaluate trading performance.

To reduce this mismatch, increasing attention has been given to execution-aware target construction. Rather than assigning labels solely according to future price direction, execution-aware approaches define trade outcomes using predefined execution rules, such as stop-loss, take-profit, and holding-period constraints. Representative examples include stop-loss-adjusted labels and the triple-barrier method proposed by \citet{lopez2018}, together with subsequent extensions that incorporate execution constraints into financial machine learning models \citep{hwang2023,fu2024}. These approaches better align the learning objective with practical trading behaviour by explicitly considering how trades would be executed rather than simply whether prices eventually move in the expected direction. Consequently, label construction has become an increasingly important component of financial machine learning systems rather than merely a preprocessing step.

Nevertheless, execution-aware learning alone does not fully bridge the gap between prediction and practical investment decision support. Although improved label construction enables models to learn trading outcomes more realistically, many studies still stop at predicting class labels or probabilities. In practice, traders require considerably richer information, including trade direction, entry price, stop-loss, take-profit, confidence level, and risk--reward ratio, all of which must be derived from the predictive model in a consistent and interpretable manner. Previous studies have also shown that trading performance is strongly influenced by exit strategies, with stop-loss and take-profit rules playing an important role even when the underlying prediction model remains unchanged \citep{vezeris2018}. Taken together, the evidence indicates that prediction and execution should be treated as closely connected components of the same decision-making process rather than as separate problems.

Overall, recent research has significantly improved the alignment between financial prediction and trading execution through execution-aware learning. However, relatively little work has integrated execution-aware prediction with structured decision generation within a unified framework. As a result, there remains a gap between models that predict market behaviour and systems that provide directly actionable investment recommendations under realistic trading constraints.

\subsection{Robust Financial Machine Learning Systems}

Beyond predictive modelling, the practical usefulness of financial machine learning systems depends heavily on methodological robustness and reproducibility. Unlike many conventional machine learning applications, financial data exhibit strong temporal dependence, distribution shifts, and frequent structural changes, making experimental design a critical factor in model evaluation. Consequently, rigorous validation strategies have become an essential component of reliable financial machine learning. Time-aware evaluation protocols are generally preferred over random train-test splits because they better reflect realistic forecasting scenarios and reduce the risk of information leakage \citep{tashman2000,bergmeir2012}. Accordingly, improvements in predictive performance should always be interpreted alongside the reliability of the underlying evaluation methodology.

Another important aspect concerns feature representation and feature-space management. Modern financial machine learning systems often generate large numbers of engineered features to capture different aspects of market behaviour. While richer feature representations can improve predictive capability, they also increase the risk of redundancy, instability, and overfitting if feature complexity is not adequately controlled. Various feature-selection techniques and dimensionality-reduction methods have therefore been proposed to improve model robustness and computational efficiency \citep{chandrashekar2014,christ2018,meinshausen2010,jolliffe2002}. Collectively, the evidence indicates that robust financial prediction depends not only on advanced learning algorithms but also on disciplined feature engineering and careful control of feature-space complexity.

The reliability of financial machine learning systems is further influenced by data quality and operational reproducibility. Financial datasets collected from different providers may contain inconsistencies, missing values, or varying historical coverage, all of which can affect downstream model performance \citep{boritz2020}. Missing-value handling has therefore become an important preprocessing step in large-scale financial applications, as inappropriate treatment may introduce bias or distort subsequent analyses \citep{chen2024,bryzgalova2025}. More broadly, reproducibility requires preprocessing procedures, learned transformations, and trained models to be preserved and reapplied consistently during model development and operational deployment.

Finally, practical financial machine learning extends beyond offline prediction to complete operational workflows. Although many studies demonstrate promising backtesting results, comparatively few address issues such as automated workflow orchestration, artefact management, scheduled execution, scalable cloud deployment, and reproducible inference pipelines. Recent advances in AutoML and cloud-native machine learning platforms have highlighted the importance of workflow automation and reproducible execution for operational AI systems \citep{feurer2015,jonas2019}. Such engineering considerations are particularly important for financial decision-support applications, where models must be updated, deployed, and executed repeatedly under consistent operating conditions.

Overall, the literature highlights the importance of robust validation, disciplined feature engineering, reliable data management, and reproducible deployment for financial machine learning. However, these aspects are often investigated independently rather than being integrated into a unified framework. Consequently, there remains a need for end-to-end financial machine learning systems that combine robust modelling, execution-aware prediction, structured decision support, and operational deployment within a single reproducible workflow.

\subsection{Summary}

Existing studies have demonstrated the potential of machine learning for financial prediction and have explored execution-aware learning, robust validation strategies, and operational deployment from different perspectives. However, these components are typically investigated in isolation rather than as parts of a unified decision-support pipeline. In particular, relatively little work integrates execution-aware label construction, multi-asset representation, feature refinement, ensemble learning, structured decision generation, and operational deployment within a single end-to-end framework. The resulting research gap provides the motivation for the framework proposed in this paper.

\section{The Proposed Framework}

\subsection{Framework Overview}

We propose an end-to-end machine learning framework for structured investment decision support. Rather than treating financial forecasting as an isolated prediction task, the proposed architecture unifies data preparation, feature engineering, execution-aware label construction, feature refinement, predictive modelling, decision generation, and operational deployment. Its purpose is to produce actionable trading information instead of standalone statistical predictions.

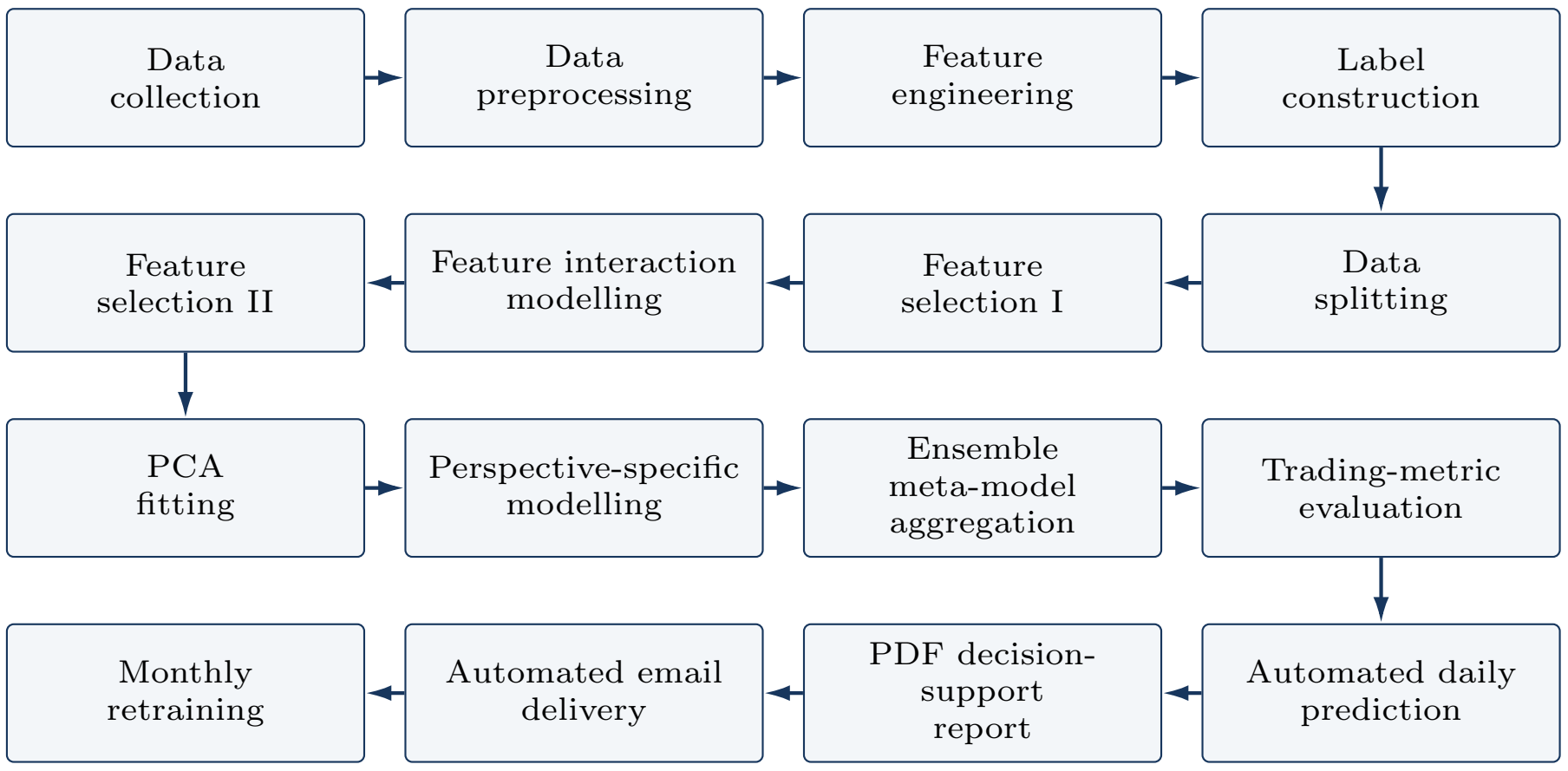
\begin{figure}[h]
    \centering
    \begin{tikzpicture}[
        stage/.style={
            draw=figureblue,
            fill=figurefill,
            rounded corners=1.5pt,
            align=center,
            text width=0.195\linewidth,
            minimum height=10.5mm,
            inner sep=2.5pt,
            font=\footnotesize\rmfamily
        },
        flow/.style={
            -{Latex[length=2mm,width=1.2mm]},
            draw=figureblue,
            line width=0.8pt
        }
    ]

    \node[stage] (s1) {Data\\collection};
    \node[stage, right=3mm of s1] (s2) {Data\\preprocessing};
    \node[stage, right=3mm of s2] (s3) {Feature\\engineering};
    \node[stage, right=3mm of s3] (s4) {Label\\construction};

    
    \node[stage, below=5mm of s4] (s5) {Data\\splitting};
    \node[stage, left=3mm of s5] (s6) {Feature\\selection I};
    \node[stage, left=3mm of s6] (s7) {Feature interaction\\modelling};
    \node[stage, left=3mm of s7] (s8) {Feature\\selection II};

    \node[stage, below=5mm of s8] (s9) {PCA\\fitting};
    \node[stage, right=3mm of s9] (s10) {Perspective-specific\\modelling};
    \node[stage, right=3mm of s10] (s11) {Ensemble meta-model\\aggregation};
    \node[stage, right=3mm of s11] (s12) {Trading-metric\\evaluation};

    \node[stage, below=5mm of s12] (s13) {Automated daily\\prediction};
    \node[stage, left=3mm of s13] (s14) {PDF decision-support\\report};
    \node[stage, left=3mm of s14] (s15) {Automated email\\delivery};
    \node[stage, left=3mm of s15] (s16) {Monthly\\retraining};

    \draw[flow] (s1.east) -- (s2.west);
    \draw[flow] (s2.east) -- (s3.west);
    \draw[flow] (s3.east) -- (s4.west);

    \draw[flow] (s4.south) -- (s5.north);
    \draw[flow] (s5.west) -- (s6.east);
    \draw[flow] (s6.west) -- (s7.east);
    \draw[flow] (s7.west) -- (s8.east);

    \draw[flow] (s8.south) -- (s9.north);
    \draw[flow] (s9.east) -- (s10.west);
    \draw[flow] (s10.east) -- (s11.west);
    \draw[flow] (s11.east) -- (s12.west);

    \draw[flow] (s12.south) -- (s13.north);
    \draw[flow] (s13.west) -- (s14.east);
    \draw[flow] (s14.west) -- (s15.east);
    \draw[flow] (s15.west) -- (s16.east);

    \end{tikzpicture}
    \caption{Overview of the proposed end-to-end framework, including data preparation, feature engineering, execution-aware label construction, feature refinement, predictive modelling, ensemble learning, and automated decision-support generation.}
    \label{fig:framework}
\end{figure}

Figure~\ref{fig:framework} illustrates the overall workflow of the proposed system. The pipeline begins by collecting daily OHLCV (open, high, low, close, and volume) data from both the target asset and a set of related financial assets. After data cleaning and temporal alignment, a large collection of technical indicators is generated to represent complementary market characteristics, including trend, momentum, and oscillation. Execution-aware labels are then constructed according to predefined stop-loss and take-profit rules, allowing the learning objective to better reflect practical trading outcomes.

The engineered feature space is subsequently refined through two rounds of feature selection with interaction modelling between them, followed by dimensionality reduction using principal component analysis (PCA). The refined representations are used to train specialist predictive models across different market perspectives. Their outputs are then combined through an ensemble learning strategy to produce class probabilities representing \textbf{buy}, \textbf{hold (no trade)}, and \textbf{sell} actions.

Instead of directly exposing these probabilities to end users, the framework converts them into structured trading recommendations containing the trade direction, entry price, stop-loss, take-profit, risk-reward ratio, and associated confidence level. The resulting outputs are automatically incorporated into a daily decision-support report, which can be generated and delivered through a cloud-based deployment pipeline.

Formally, let $\mathbf{x}_t$ denote the feature vector constructed from all market information available at time $t$, including both the target asset and related assets. The predictive model learns the mapping
\begin{equation}
f:\mathbf{x}_t \rightarrow \mathbf{p}_t,
\end{equation}
where $\mathbf{p}_t$ represents the predicted probability distribution over the possible trading actions. A deterministic post-processing procedure subsequently transforms the predicted probabilities into structured trading information that can be incorporated directly into the decision-support workflow.

Overall, the proposed framework extends conventional financial prediction pipelines by integrating predictive modelling with execution-aware decision generation and operational deployment within a unified architecture.

\subsection{Multi-Asset Data Construction}

\subsubsection{Data Collection and Representation}

A multi-asset representation provides broader market context for predicting the target asset. GBP/USD serves as the prediction target, while the auxiliary market universe contains 253 related financial instruments, including equities, foreign exchange pairs, commodities, metals, energy products, market indices, exchange-traded funds, and sector-specific instruments. Movements in these related markets may provide useful contextual information beyond the historical behaviour of GBP/USD alone.

The raw market history spans the period from early 2004 to late April 2026. Daily OHLCV data are collected for each asset from Yahoo Finance through the \texttt{yfinance} Python interface. Table~\ref{tab:data-example} provides an example of the collected target-asset data together with the aligned cross-asset representation used by the proposed framework.

\begin{table*}[h]
\centering
\caption{Example of raw daily OHLCV data for the target asset (GBP/USD) and the aligned cross-asset representation used for feature engineering.}
\label{tab:data-example}

\resizebox{\textwidth}{!}{%
\begin{tabular}{@{}lrrrrr|rrrrrrrrr@{}}
\toprule
& \multicolumn{5}{c|}{\textbf{Target asset (GBP/USD)}}
& \multicolumn{9}{c}{\textbf{Aligned cross-asset representation}} \\
\cmidrule(lr){2-6}
\cmidrule(lr){7-15}

Date
& Open & High & Low & Close & Volume
& SP500 Close & SP500 Vol
& Gold Close & Gold Vol
& DXY Close & Oil Close
& US10Y Yield & BTC Close & BTC Vol \\
\midrule

2004-01-03
& 1.7795 & 1.7852 & 1.7721 & 1.7818 & 0
& 1118.9 & 1.32M & 415.2 & 82K
& 85.6 & 33.8 & 4.21 & NaN & NaN \\

2004-01-04
& 1.7816 & 1.7903 & 1.7775 & 1.7884 & 0
& 1122.4 & 1.41M & 417.1 & 91K
& 85.8 & 34.1 & 4.25 & NaN & NaN \\

2004-01-05
& 1.7882 & 1.7930 & 1.7812 & 1.7841 & 0
& 1115.7 & 1.38M & 414.6 & 88K
& 86.2 & 33.5 & 4.19 & NaN & NaN \\

\multicolumn{15}{c}{\ldots} \\

2026-04-23
& 1.2638 & 1.2714 & 1.2582 & 1.2697 & 0
& 5124.7 & 3.84M & 2381.3 & 210K
& 103.8 & 84.5 & 3.96 & 74200 & 21K \\

2026-04-24
& 1.2696 & 1.2758 & 1.2633 & 1.2721 & 0
& 5148.1 & 3.92M & 2394.7 & 198K
& 103.2 & 85.2 & 3.91 & 74850 & 19K \\

2026-04-27
& 1.2723 & 1.2785 & 1.2674 & 1.2749 & 0
& 5162.3 & 4.01M & 2408.2 & 225K
& 102.7 & 86.1 & 3.88 & 75520 & 23K \\

\bottomrule
\end{tabular}%
}
\end{table*}

Two important data-quality characteristics are evident in Table~\ref{tab:data-example}. First, GBP/USD volume is recorded as zero because spot foreign exchange does not have a centralised exchange that reports trading volume in the same manner as equities. Second, Bitcoin observations are unavailable for the earliest dates because historical data are unavailable before the asset began trading. These characteristics illustrate two common challenges encountered in financial datasets: non-informative volume values and temporally incomplete observations. They motivate the preprocessing procedures described in the following subsection.

More generally, financial time series differ from many conventional machine learning datasets because they exhibit strong temporal dependence, substantial noise, missing observations, and changing statistical properties. Consequently, reliable preprocessing, including temporal alignment, missing-value handling, and data cleaning, is essential before feature engineering and model training.

\subsubsection{Data Cleaning and Temporal Alignment}

The collected datasets are processed through a unified preprocessing pipeline. First, date formats and column names are standardised across all assets, and only the required OHLCV variables are retained. Assets with incomplete downloads or empty records are discarded automatically. 
Next, all auxiliary assets are temporally aligned with the target asset so that every observation corresponds to a common trading timeline. Alignment to a common timeline ensures that every feature available at a prediction point originates from information that would have been observable at that time. Missing values are then handled using a deterministic procedure. Forward filling is applied as the primary imputation strategy to preserve temporal continuity, while backward filling is used only for remaining isolated missing values. Columns or rows containing no valid observations are removed from the dataset. For assets for which trading volume is not meaningfully reported, such as certain foreign exchange instruments, zero-volume entries are treated as missing values rather than genuine observations.

After preprocessing, the target-asset dataset contains 5,814 daily observations, while the aligned cross-asset representation contains 1,138 raw input variables. The final dataset is constructed by horizontally concatenating the aligned OHLCV variables across all selected assets. Rather than explicitly modelling pairwise relationships during preprocessing, subsequent feature-engineering stages and predictive models learn them automatically from the aligned multi-asset representation.

The entire preprocessing stage is configuration-driven, allowing the asset universe, temporal range, and preprocessing parameters to be modified without changing the implementation logic. Configuration-driven preprocessing therefore facilitates reuse across different assets and experimental settings while preserving a consistent data-processing workflow.

\subsection{Feature Engineering}

To capture market characteristics that cannot be represented by raw prices alone, the aligned multi-asset data are transformed into a high-dimensional feature representation. Instead of relying solely on raw prices or a small set of manually selected indicators, the pipeline generates a diverse collection of technical features describing trend, momentum, volatility, oscillation, channel behaviour, and candlestick patterns. The resulting representation enables the predictive models to exploit different aspects of market dynamics.

Feature generation is implemented using vectorised \texttt{pandas} and \texttt{NumPy} operations. All features are computed exclusively from historical observations, ensuring that future information is not introduced during feature construction. A parameter-driven generation strategy applies predefined feature templates across multiple settings to produce a comprehensive feature space automatically.

The generated features can be broadly grouped into three complementary perspectives:

\begin{itemize}
\item \textbf{Trend features}, which describe long-term market direction and persistence.
\item \textbf{Momentum features}, which characterise the speed and strength of recent price movements.
\item \textbf{Oscillation features}, which capture cyclic behaviour and potential market reversals.
\end{itemize}

Within these categories, the framework generates a wide range of commonly used technical indicators, including moving averages, the relative strength index (RSI), Bollinger Bands, rolling volatility statistics, channel-based indicators, and candlestick-derived features. Feature generation is performed for both the target asset and the related-asset universe.

Before feature refinement, the framework produces a total of 34,982 candidate features, comprising 321 features derived from the target asset and 34,661 features extracted from the related assets. Deliberate over-completeness allows subsequent feature-selection stages to identify the most informative variables while reducing redundancy and noise.

To ensure reproducibility during deployment, the framework stores the final retained feature definitions after model training. During daily inference, the retained feature names are automatically parsed to reconstruct the corresponding features using the latest market data, ensuring identical feature definitions during training and deployment.


\subsection{Execution-Aware Label Construction}

Unlike conventional financial prediction methods that formulate the problem as directional forecasting, the proposed framework constructs execution-aware labels that explicitly consider practical trading outcomes. Instead of asking only whether the future price moves upward or downward, the framework evaluates whether a predefined trading strategy would successfully reach its profit target before triggering its stop-loss. Consequently, the learning objective is aligned more closely with practical trading decisions.

\begin{figure}
    \centering
    \includegraphics[
        width=\linewidth
    ]{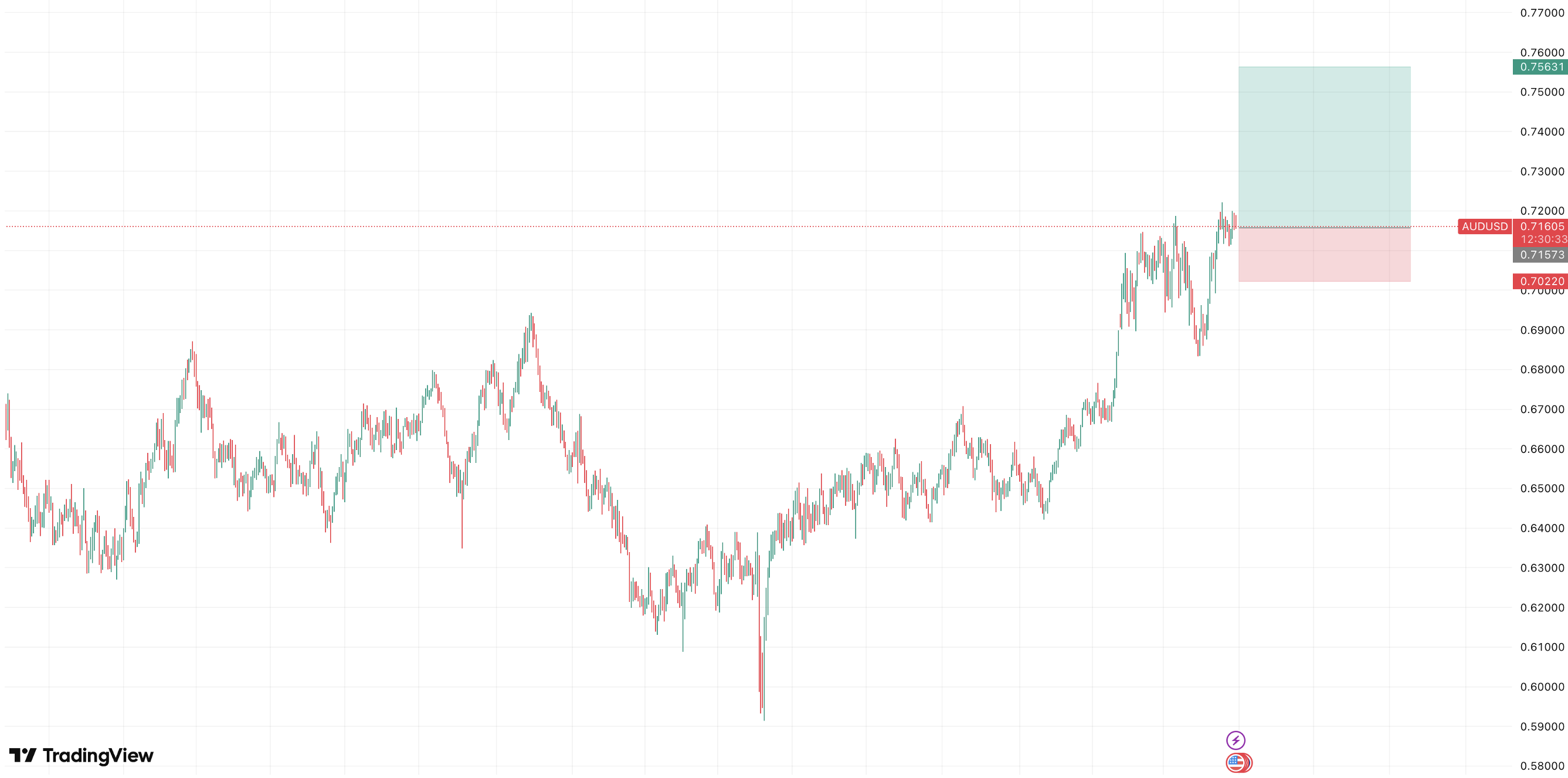}
    \caption{TradingView screenshot illustrating execution-aware labelling using ATR-based stop-loss and take-profit barriers. The realised price path determines the trade outcome.}
    \label{fig:execution-aware-label}
\end{figure}

Figure~\ref{fig:execution-aware-label} illustrates the execution-aware labelling strategy. Each potential trade is defined by three components: an entry price, a stop-loss level representing the maximum acceptable loss, and a take-profit level representing the desired profit target. A trade is regarded as successful only if the take-profit level is reached before the stop-loss level during the predefined evaluation horizon.

For every trading day, the framework assumes that a hypothetical trade is opened at the closing price. The Average True Range (ATR) is first computed to estimate recent market volatility and is then used to determine adaptive stop-loss and take-profit levels according to a predefined risk--reward ratio.

After the trading boundaries have been established, subsequent market prices are examined sequentially over the prediction horizon. The labelling procedure operates as follows:

\begin{itemize}
\item If the take-profit level of the long position is reached before its stop-loss, the observation receives the buy label;
\item If the stop-loss level of the long position is reached first, the corresponding short position is evaluated;
\item If the short position subsequently reaches its take-profit level before the evaluation horizon expires, the observation receives the sell label;
\item If neither position produces a successful trade before the evaluation horizon expires, the observation receives the hold (no-trade) label.
\end{itemize}

The resulting labels allow the predictive models to learn execution outcomes rather than simple price direction. Compared with conventional directional labelling, this formulation incorporates both the magnitude and temporal sequence of future price movements. Consequently, model predictions become directly associated with predefined execution rules instead of abstract directional movements. The resulting labels therefore provide a more appropriate learning objective for practical decision-support systems, where trading success depends not only on whether prices eventually move in the predicted direction, but also on whether predefined risk-management constraints are satisfied during the trade.

\subsection{Feature Refinement}

The large feature space generated in the previous stage inevitably contains redundant, noisy, and highly correlated variables. Directly training predictive models on all generated features would substantially increase computational cost and may reduce generalisation performance. Feature refinement therefore proceeds through a multi-stage pipeline comprising feature selection, interaction modelling, and dimensionality reduction.

The refinement process consists of four sequential steps: (i) first-round feature selection, (ii) feature interaction modelling, (iii) second-round feature selection, and (iv) principal component analysis (PCA).

During the first round of feature selection, three complementary categories of selection methods are employed.

The first category consists of filter-based methods, including variance filtering, mutual information, ANOVA F-statistics, Fisher score, single-feature ROC-AUC, distance correlation, temporal stability mutual information, Kruskal--Wallis statistics, and a trading-specific expectancy score. Raw OHLCV variables are always retained because they provide the fundamental price information on which engineered indicators depend. Features selected by any filter method are preserved using a union strategy.

The retained features are subsequently refined using wrapper-based approaches, including recursive feature elimination with cross-validation (RFECV), Boruta, and a temporally stable feature-selection strategy. Again, the union of all selected features is retained.

Finally, embedded feature-selection methods are applied, including LightGBM feature importance, XGBoost SHAP importance, and Elastic Net logistic regression coefficients. Their outputs are similarly combined using a union rule to obtain the final first-round feature subset.

After the first selection stage, second-order polynomial interaction features are generated using only the retained variables. Restricting interaction modelling to the selected subset avoids the combinatorial explosion that would arise from applying polynomial expansion directly to the original feature space. Second-order interactions are used because they capture pairwise nonlinear relationships while remaining computationally tractable.

A second round of feature selection is then performed using the same filter, wrapper, and embedded strategies to remove weak or redundant interaction features. Instead of adopting a fixed global threshold, the number of retained features is determined according to the size of each feature group, allowing the refinement process to remain flexible across different datasets.

Before PCA, each feature group is partitioned chronologically within every walk-forward fold into training, validation, and test sets. PCA is then fitted exclusively on the training data, and the fitted transformation is reused without modification on the corresponding validation, test, and deployment datasets. An identical training-only rule governs both rounds of feature selection and interaction modelling, preventing information leakage and preserving consistency between model development and operational deployment.

All retained feature lists, interaction definitions, feature-selection masks, and fitted PCA transformations are stored automatically after training and reused during subsequent inference, providing a reproducible and deployment-ready feature-refinement pipeline.

\subsection{Predictive Models and Ensemble Learning}
Following feature refinement, predictive modelling is designed to capture complementary aspects of market behaviour while reducing dependence on any individual learning algorithm. To this end, models are trained independently on six complementary feature groups: target-asset trend, related-asset trend, target-asset momentum, related-asset momentum, target-asset oscillation, and related-asset oscillation.

The modelling framework includes fourteen representative machine learning algorithms covering linear models, tree-based methods, kernel methods, probabilistic models, neural networks, and hidden-state models. Specifically, the model library consists of XGBoost, CatBoost, random forest, extra trees, logistic regression, elastic-net logistic regression, a linear support vector machine (SVM), a radial basis function (RBF) SVM, gradient-boosted logistic regression, k-nearest neighbours, Gaussian naive Bayes, AdaBoost, a multilayer perceptron, and hidden Markov models.

To improve adaptability to changing market conditions, most predictive models incorporate time-decay weighting during training, allowing more recent observations to contribute more strongly than older historical samples.

Training is performed independently for each feature group across the six walk-forward folds. Within every fold, both untuned and hyperparameter-optimised versions of each model are trained. Hyperparameter optimisation and selection between the tuned and untuned versions are conducted exclusively on the validation partition. The version achieving the higher validation-period Sharpe ratio is retained as the representative model for that fold, while the test partition remains unseen until final backtesting.

Across six feature groups, fourteen model families, and six walk-forward folds, the framework retains a total of $6 \times 14 \times 6 = 504$ fold-level predictive models.


To construct the ensemble, the model with the strongest validation performance is selected from the six fold-level candidates for each combination of feature group and model family, resulting in 84 base learners. Each base learner outputs class probabilities corresponding to the three trading actions: buy, hold (no trade), and sell. The resulting probability estimates are concatenated into a 252-dimensional meta-feature vector, which serves as the input to the final ensemble model.

Rather than combining predictions through simple averaging or majority voting, the proposed framework adopts stacked ensemble learning, allowing the meta-model to learn how individual base learners complement one another under different market conditions. This hierarchical architecture exploits the diversity of both feature representations and predictive models, enabling the final ensemble to produce more robust trading decisions than any individual model alone.

All trained models, probability outputs, ensemble parameters, and configuration files are automatically stored after training. These artefacts are subsequently reused during operational deployment, ensuring that identical predictive models are applied during daily inference without retraining or manual intervention.

\subsection{Decision-Support Generation and Deployment}

The final stage of the proposed framework transforms predictive probabilities into structured investment recommendations and deploys the complete workflow as an automated decision-support system. Rather than presenting only the predicted class probabilities, the framework converts model outputs into interpretable trading information that can be reviewed directly by users.

Following ensemble prediction, the model produces posterior probabilities for the three trading actions: buy, hold (no trade), and sell. A deterministic post-processing procedure then converts the probabilities into a structured trade recommendation. The generated recommendation includes the trade direction, entry price, ATR-based stop-loss, ATR-based take-profit, risk--reward ratio, confidence level, and probability estimate for each trading action.

To ensure consistency between model training and operational deployment, the same ATR-based execution rules used during execution-aware label construction are reused during report generation. The latest completed daily market data are first retrieved, after which the ATR is computed to derive the corresponding stop-loss and take-profit levels. Consequently, the reported trading recommendations remain fully consistent with the execution assumptions adopted during model development.

The final recommendation is automatically converted into a structured PDF report using the ReportLab library. The report summarises the predicted trading signal and associated execution parameters, thereby providing actionable decision-support information instead of standalone model predictions.

To support operational deployment, the entire workflow is implemented on Google Cloud Platform (GCP). Vertex AI Pipelines orchestrate the end-to-end execution process, while Google Cloud Storage (GCS) stores trained models, intermediate artefacts, configuration files, and generated reports. Docker images are managed through Artifact Registry to ensure reproducible execution across different computing environments.

Daily report delivery is fully automated using SendGrid. After each prediction cycle, the generated PDF report is retrieved from cloud storage and delivered automatically to the predefined recipients via email. Monthly model retraining and daily prediction are scheduled independently, allowing computationally intensive retraining to be performed once per month while lightweight prediction and report generation are executed on every trading day.

Overall, the deployment architecture extends the proposed framework beyond offline experimentation by integrating predictive modelling, report generation, cloud orchestration, and automated delivery into a complete operational decision-support pipeline.

\subsection{Implementation Details}

Implementation relies primarily on Python. Data processing and numerical computation are performed using \texttt{pandas} and \texttt{NumPy}, while \texttt{scikit-learn} provides preprocessing, feature selection, dimensionality reduction, and baseline machine learning algorithms. Market data are collected through the \texttt{yfinance} interface, and additional market analysis is supported using TradingView and TradeZella during system development.

Model development and debugging are conducted in Jupyter notebooks, whereas operational deployment runs on GCP. Intermediate datasets are stored in CSV and Parquet formats, while trained models and preprocessing objects are serialised using the Python \texttt{pickle} module. Persistent cloud artefacts, including trained models, configuration files, generated reports, and execution records, are maintained within Google Cloud Storage.

To improve reproducibility, all major components of the framework are configuration-driven. Model parameters, feature-generation settings, selected assets, validation folds, report options, and deployment settings are specified externally rather than being embedded within the source code. This allows the same implementation to be reused across different assets and experimental settings without modifying the underlying workflow.

Throughout the pipeline, intermediate artefacts generated during training, including retained feature lists, fitted preprocessing objects, PCA transformations, trained predictive models, and ensemble parameters, are automatically saved and subsequently reused during deployment. This ensures identical transformations during both model development and operational inference while reducing the risk of inconsistencies between offline experiments and daily prediction.

The implementation adopts a modular software architecture in which data preparation, feature engineering, feature refinement, predictive modelling, report generation, and deployment are implemented as independent but connected components. Such modularity facilitates maintenance, extensibility, reproducibility, and future integration of additional predictive models or deployment environments while preserving a consistent end-to-end workflow. Having described the architecture and implementation, the following section evaluates its predictive, trading, and operational performance.

\section{Results and Discussion}

\subsection{Evaluation Protocol}

The proposed framework is evaluated using a six-fold walk-forward design in which models are repeatedly trained on earlier historical data and tested on later unseen periods. This design is more appropriate for financial time-series modelling than random splitting because it preserves temporal order and reduces the risk of information leakage from future observations into the training process.

The evaluation is also month-specialised. Instead of testing on a single continuous future block, the framework evaluates performance across repeated calendar-month conditions. This provides a more structured assessment of whether the framework performs consistently under recurring seasonal market conditions.

\begin{figure}[h]
    \centering
    \includegraphics[
        width=\linewidth,
        height=0.46\textheight,
        keepaspectratio
    ]{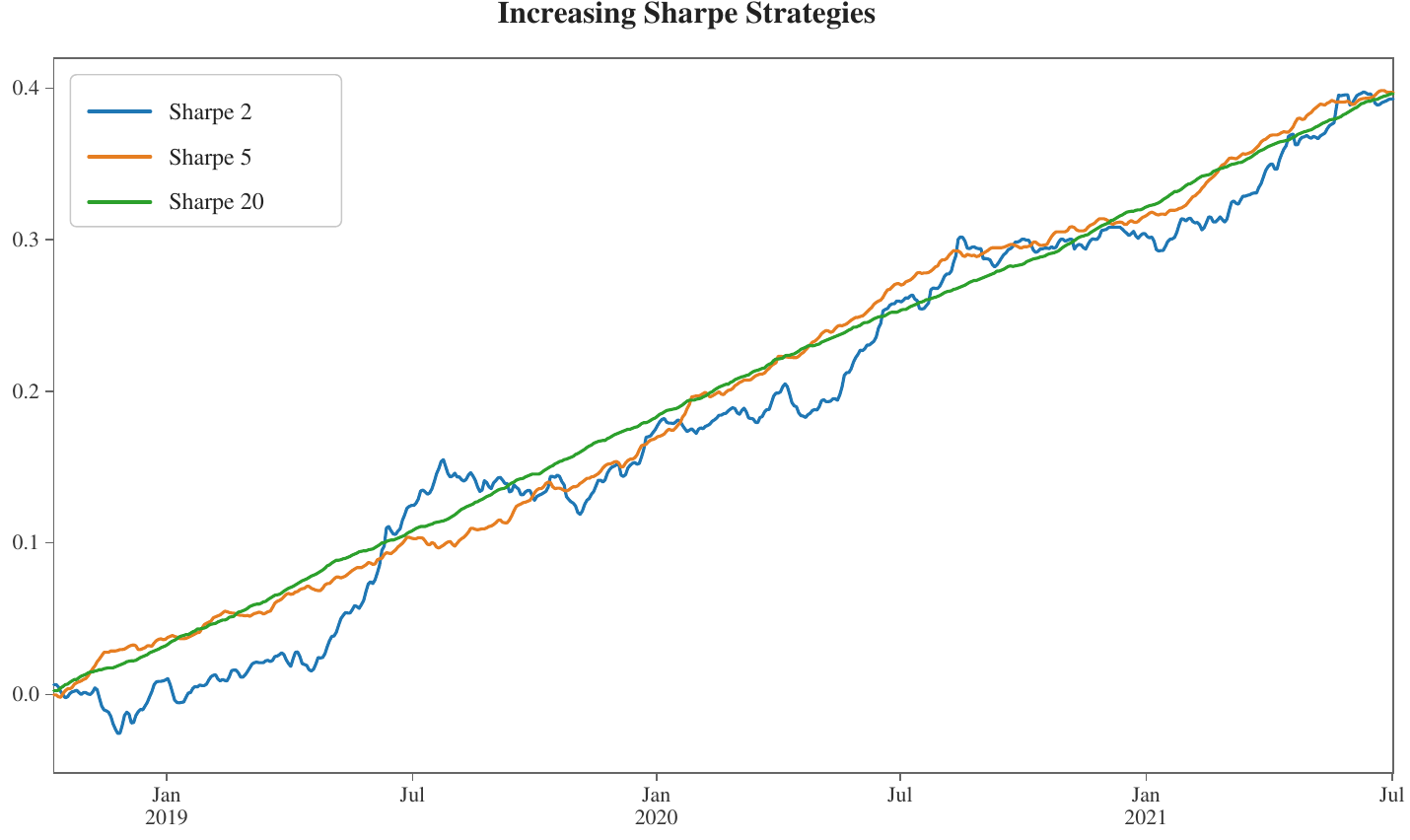}
    \caption{Illustration of how higher-Sharpe strategies produce smoother and more stable returns, whereas lower-Sharpe strategies exhibit greater fluctuations despite similar overall growth.}
    \label{fig:sharpe-ratio-illustration}
\end{figure}

Figure~\ref{fig:sharpe-ratio-illustration} provides a visual interpretation of the primary risk-adjusted metric used in the evaluation. Although the illustrated strategies achieve similar cumulative growth, the higher-Sharpe paths are smoother and exhibit smaller fluctuations. The comparison therefore demonstrates that the Sharpe ratio captures return consistency rather than final return alone.

Performance is evaluated primarily using the Sharpe ratio, maximum drawdown, and profit factor. The Sharpe ratio measures risk-adjusted return, maximum drawdown measures the largest peak-to-trough decline, and profit factor measures the ratio of total profits to total losses. Supporting metrics, including win rate, expectancy, and average R-multiple, are also reported to provide additional insight into trading behaviour.

All learned preprocessing and transformation steps are fitted only on the training partition of each fold and then reused unchanged on validation, test, and later prediction data. The restriction covers feature selection, interaction modelling, PCA, and model training. Hyperparameter tuning is also performed within the same time-safe fold structure. The validation data are used for model selection, while the test data remain unseen until the final backtesting stage.

Testing is performed through sequential out-of-sample backtesting. For each retained fold-level model, predictions are generated in chronological order and converted into trading signals, where 1 denotes a buy signal, $-1$ denotes a sell signal, and 0 denotes no trade. The chronologically generated signals are then passed to a backtesting engine using the observed OHLC market data. The engine applies the same ATR-based stop-loss and take-profit rules used in the label-construction stage. Therefore, models are evaluated not only by whether they predict the correct class, but also by whether their predictions would generate useful trades under predefined execution rules.

Three comparisons are used to evaluate the principal design choices. The first compares execution-aware labels with conventional directional labels. The second compares the complete integrated pipeline with a simpler baseline that applies one non-specialised model directly to the engineered feature set. The third compares the multi-asset, month-specialised configuration with a single-asset configuration using a single continuous future hold-out block, referred to as weaker validation. Wherever possible, the remaining pipeline components are held constant so that each comparison reflects the stated design difference.

\subsection{Execution-Aware Labelling versus Directional Labelling}

Table~\ref{tab:execution-aware-comparison} compares the proposed execution-aware label construction with a conventional Directional labelling baseline. The execution-aware version outperforms the directional baseline across all three primary metrics. The Sharpe ratio increases from 0.92 to 1.47, maximum drawdown improves from $-23.8\%$ to $-15.6\%$, and profit factor increases from 1.28 to 1.74.

\begin{table}[h]
    \centering
    \standardtablestyle
    \caption{Comparison of execution-aware labelling with directional labelling under forward testing.}
    \begin{tabular}{@{}>{\raggedright\arraybackslash}p{0.25\linewidth}cccccc@{}}
        \toprule
        Configuration &
        \shortstack{Sharpe\\ratio} &
        \shortstack{Maximum\\drawdown} &
        \shortstack{Profit\\factor} &
        \shortstack{Win\\rate} &
        \shortstack{Expectancy\\(R)} &
        \shortstack{Average\\R-multiple} \\
        \midrule
        Directional-labelling baseline
        & 0.92 & $-23.8\%$ & 1.28 & 51.4\% & 0.18 & 0.62 \\

        Execution-aware labelling
        & 1.47 & $-15.6\%$ & 1.74 & 48.9\% & 0.41 & 1.12 \\
        \midrule

        Difference
        & $+0.55$
        & $+8.2\,\mathrm{pp}$
        & $+0.46$
        & $-2.5\,\mathrm{pp}$
        & $+0.23$
        & $+0.50$ \\
        \bottomrule
    \end{tabular}
    \label{tab:execution-aware-comparison}
\end{table}

These results indicate that aligning the learning target with trade execution rules improves the practical usefulness of the model. The directional baseline predicts only whether prices move up or down, whereas the execution-aware version is trained on outcomes that already reflect stop-loss and take-profit conditions. Consequently, the training objective is more consistent with the way trades are evaluated in practice.

The supporting metrics reveal an important trade-off. Win rate decreases slightly from 51.4\% to 48.9\%, but expectancy increases from 0.18 to 0.41 and average R-multiple increases from 0.62 to 1.12. Despite the lower win rate, the higher expectancy and larger average R-multiple indicate that execution-aware labelling selects fewer but more valuable trades and produces stronger risk-adjusted outcomes.

\subsection{Integrated Pipeline versus Single-Model Baseline}

Table~\ref{tab:integrated-pipeline-baseline} compares the full integrated framework with a simpler single-model baseline. The integrated pipeline achieves a Sharpe ratio of 1.72, compared with 1.03 for the baseline. Maximum drawdown improves from $-22.4\%$ to $-13.8\%$, while profit factor increases from 1.36 to 1.98.

\begin{table}[h]
    \centering
    \standardtablestyle
    \caption{Comparison of the integrated pipeline with the simpler single-model baseline.}
    \begin{tabular}{@{}>{\raggedright\arraybackslash}p{0.29\linewidth}ccccc@{}}
        \toprule
        Configuration &
        \shortstack{Sharpe\\ratio} &
        \shortstack{Maximum\\drawdown} &
        \shortstack{Profit\\factor} &
        \shortstack{Win\\rate} &
        \shortstack{Expectancy\\(R)} \\
        \midrule
        Simpler single-model baseline
        & 1.03 & $-22.4\%$ & 1.36 & 52.1\% & 0.22 \\

        Integrated pipeline
        & 1.72 & $-13.8\%$ & 1.98 & 49.3\% & 0.48 \\
        \midrule

        Difference
        & $+0.69$
        & $+8.6\,\mathrm{pp}$
        & $+0.62$
        & $-2.8\,\mathrm{pp}$
        & $+0.26$ \\
        \bottomrule
    \end{tabular}
    \label{tab:integrated-pipeline-baseline}
\end{table}

These improvements suggest that the proposed system-level design adds value beyond a single predictive model. The baseline uses one non-specialised model directly on the engineered feature set, whereas the proposed framework combines feature refinement, specialist modelling, and ensemble aggregation. The stronger performance indicates that different modelling components can capture complementary market information and produce a more reliable final signal when combined within a structured pipeline.

The supporting metrics again show that improvement does not come from a higher win rate. Win rate decreases from 52.1\% to 49.3\%, while expectancy increases from 0.22 to 0.48. The combination of a lower win rate and higher expectancy indicates that the integrated pipeline produces fewer but more effective trading opportunities. However, the result should still be interpreted within the tested setting, and the magnitude of improvement should not be assumed to generalise unchanged to all future market conditions.

\subsection{Multi-Asset Representation and Temporal Validation}

Table~\ref{tab:combined-comparison} evaluates the combined effect of multi-asset representation and month-specialised temporal validation. The single-asset configuration with weaker validation achieves a Sharpe ratio of 1.18, maximum drawdown of $-19.7\%$, and profit factor of 1.49. In comparison, the multi-asset configuration with month-specialised validation achieves a Sharpe ratio of 1.64, maximum drawdown of $-14.2\%$, and profit factor of 1.91.

\begin{table}[h]
    \centering
    \standardtablestyle
    \caption{Comparison of the single-asset configuration with weaker validation and the multi-asset configuration with month-specialised validation.}
    \begin{tabular}{@{}>{\raggedright\arraybackslash}p{0.34\linewidth}ccccc@{}}
        \toprule
        Configuration &
        \shortstack{Sharpe\\ratio} &
        \shortstack{Maximum\\drawdown} &
        \shortstack{Profit\\factor} &
        \shortstack{Win\\rate} &
        \shortstack{Expectancy\\(R)} \\
        \midrule
        Single-asset with weaker validation
        & 1.18 & $-19.7\%$ & 1.49 & 50.8\% & 0.29 \\

        Multi-asset with month-specialised validation
        & 1.64 & $-14.2\%$ & 1.91 & 49.6\% & 0.47 \\
        \midrule

        Difference
        & $+0.46$
        & $+5.5\,\mathrm{pp}$
        & $+0.42$
        & $-1.2\,\mathrm{pp}$
        & $+0.18$ \\
        \bottomrule
    \end{tabular}
    \label{tab:combined-comparison}
\end{table}

The comparison indicates that the combined use of broader market context and stronger temporal validation improves forward-testing performance. The multi-asset component supplies information from related markets, while month-specialised validation provides a stricter time-aware evaluation setting across repeated calendar-month conditions. Because both factors change simultaneously, the observed improvement should be attributed to the combined configuration rather than to either component individually.

Improvements are consistent across the primary metrics. Sharpe ratio increases by 0.46, maximum drawdown improves by 5.5 percentage points, and profit factor increases by 0.42. Expectancy also rises from 0.29 to 0.47, indicating stronger average trade outcomes. The win rate decreases slightly from 50.8\% to 49.6\%, again suggesting that the stronger configuration improves trade quality rather than simply increasing the number of successful trades.


\subsection{Operational Validation and Deployment}

\begin{figure}[h]
    \centering
    \includegraphics[
        width=\linewidth,
        height=0.30\textheight,
        keepaspectratio
    ]{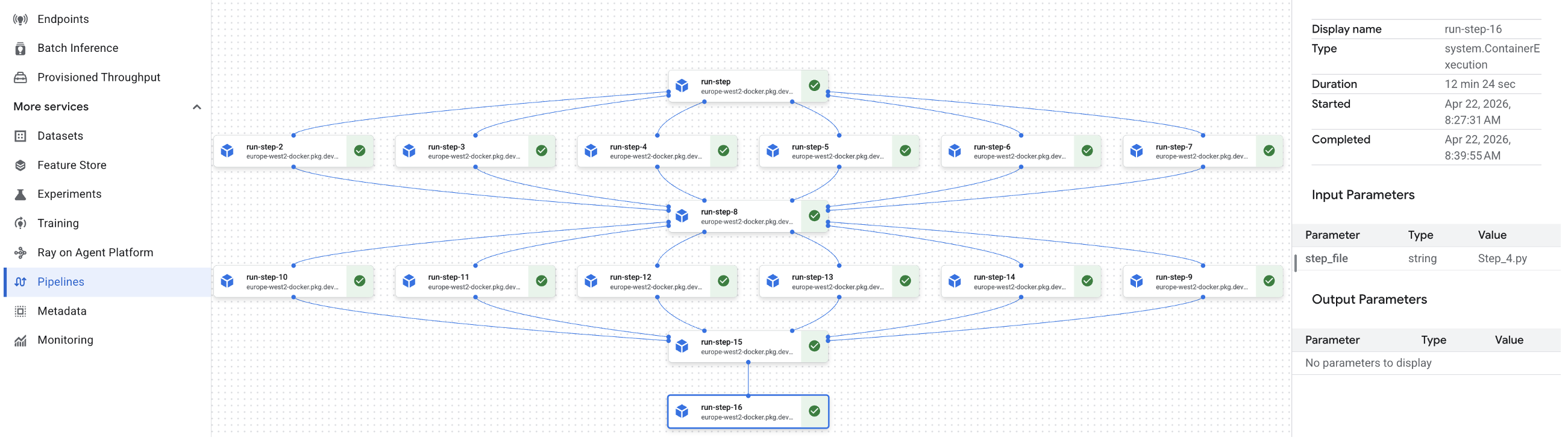}
    \caption{Vertex AI console showing a successful pipeline run. The completed stages and run metadata document execution under the evaluated conditions.}
    \label{fig:vertex-ai-pipeline}
\end{figure}

Operational validation assesses whether the proposed system functions beyond local development and historical backtesting. Figure~\ref{fig:vertex-ai-pipeline} documents a successful Vertex AI pipeline run. The completed pipeline stages and recorded run metadata provide evidence of successful execution in a managed cloud environment.

\begin{figure}
    \centering
    \includegraphics[
        width=0.6\linewidth
    ]{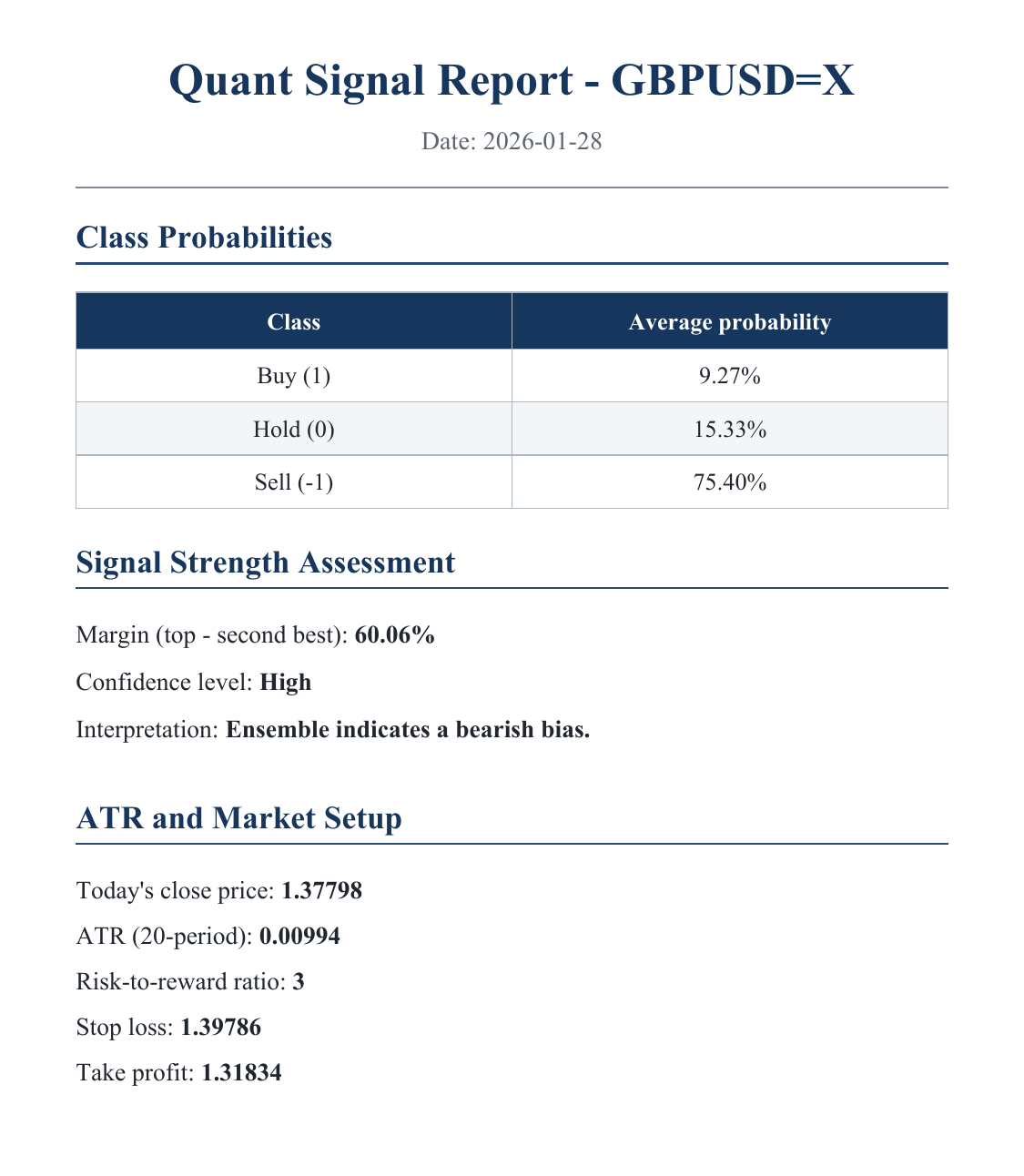}
    \caption{Example of an automatically generated decision-support report showing class probabilities, signal confidence, and ATR-based trade levels.}
    \label{fig:sample-report}
\end{figure}

Figure~\ref{fig:sample-report} demonstrates the final decision-support artefact produced by the pipeline. The example report presents the predicted class probabilities, identifies the dominant sell signal with high confidence, and translates that prediction into ATR-based stop-loss and take-profit levels. The figure therefore shows how statistical model outputs are converted into information that can be interpreted and used by an end user.

Additionally, Table~\ref{tab:daily-automation-evidence} summarises the completion of the daily operational workflow. All four required tasks, i.e., prediction generation, PDF generation, report storage, and email delivery, were completed automatically. Together, the two figures and the table demonstrate successful execution, decision-support generation, and automated delivery under the evaluated conditions.

Cloud deployment also provides practical scalability benefits. In this implementation, execution time decreased from approximately 42 hours in a local sequential setting to approximately 8 hours in the Vertex AI pipeline, corresponding to an approximate reduction of 81\%. Beyond serving as a delivery mechanism, this reduction improves the feasibility of repeated execution.

However, these operational results provide evidence of functionality under normal operating conditions rather than exhaustive proof of long-term robustness. A successful pipeline run, a generated report, and a completed automation workflow demonstrate that the system operates correctly under the tested conditions. They do not fully evaluate behaviour under delayed market data, storage failure, dependency errors, infrastructure disruption, or email delivery failure. Broader stress testing would be required before making stronger claims about long-term operational resilience.


\FloatBarrier

\begin{table}
    \centering
    \standardtablestyle
    \caption{Evidence that the daily prediction, reporting, storage, and email delivery tasks completed automatically.}
    \begin{tabular}{@{}lc@{}}
        \toprule
        Daily task & Completed automatically? \\
        \midrule
        Daily prediction generated & Yes \\
        Daily PDF generated        & Yes \\
        Report stored              & Yes \\
        Email sent                 & Yes \\
        \bottomrule
    \end{tabular}
    \label{tab:daily-automation-evidence}
\end{table}

\subsection{Reuse Across Assets and Configurations}

Table~\ref{tab:reuse-across-assets} evaluates whether the framework can be reused across different assets and configurations without structural redesign. The pipeline ran successfully in all three test cases, i.e., the baseline asset GBP/USD, the alternative asset USD/CHF, and the alternative model configuration, including successful report generation and email delivery.

Successful completion across all three test cases supports the architectural consistency of the proposed framework. Changing only the configuration parameters allows the same pipeline structure to be reused, indicating that the implementation is not tightly coupled to a single asset or model configuration. Such reusability is important for practical decision-support systems, where maintainability and configurability are necessary for repeated operation.

Nevertheless, the test provides evidence of structural reuse rather than broad predictive generalisation. Successful execution under the tested variations does not establish that predictive performance remains equally strong across all assets or market regimes. Moreover, the evaluation includes only one alternative asset and one alternative model configuration. The evidence therefore demonstrates feasibility and modularity, but not comprehensive robustness across the broader universe of financial markets.

\begin{table}[htbp]
    \centering
    \standardtablestyle
    \caption{Evidence of successful reuse across the tested assets and model configurations.}
    \begin{tabular}{@{}>{\raggedright\arraybackslash}p{0.27\linewidth}cccc@{}}
        \toprule
        Test case &
        \shortstack{Structural redesign\\required?} &
        \shortstack{Successful\\run?} &
        \shortstack{PDF\\generated?} &
        \shortstack{Email\\sent?} \\
        \midrule
        Baseline target asset GBP/USD
        & No & Yes & Yes & Yes \\

        Alternative asset USD/CHF
        & No & Yes & Yes & Yes \\

        Alternative model configuration
        & No & Yes & Yes & Yes \\
        \bottomrule
    \end{tabular}
    \label{tab:reuse-across-assets}
\end{table}

\subsection{Discussion}

Within the evaluated setting, the experimental results indicate that integrating execution-aware learning with an end-to-end decision-support framework improves the practical usefulness of financial machine learning. Unlike conventional prediction-oriented approaches, the proposed framework is designed to align model optimisation with realistic trading execution and subsequently transform predictive outputs into structured investment recommendations. Improving practical trading systems therefore requires not only stronger predictive models but also closer alignment between model-training objectives and real trading decision processes.

The results also highlight the importance of viewing financial machine learning as a complete system rather than a collection of independent modelling components. The improvements observed in this study cannot be attributed to a single algorithm alone. Instead, they arise from interactions among multiple stages, including execution-aware label construction, multi-asset representation, feature refinement, specialist modelling, ensemble aggregation, and structured decision generation. Such a systems-oriented design enables the proposed framework to address several limitations commonly observed in prediction-oriented financial machine learning studies.

Another important observation is that practical deployment should be considered an integral component of financial machine learning research rather than an implementation detail. In many existing studies, model development ends with offline evaluation or historical backtesting. In contrast, the proposed framework demonstrates that prediction, reporting, workflow orchestration, and automated deployment can be integrated into a unified and reproducible pipeline. Such integration improves the practicality, transparency, and repeatability of machine learning systems intended for real-world decision support.

Overall, the findings support a systems-oriented view of financial machine learning in which prediction, execution, risk management, operational deployment, and human decision support are considered jointly. From this perspective, the proposed framework represents a step towards more practical and reproducible financial decision-support systems. The limitations of the current evidence and the resulting priorities for future research are considered next.

\section{Limitations and Future Work}
Although the proposed framework demonstrates encouraging results, several limitations remain. First, the empirical evaluation covers a limited range of assets, market conditions, and model configurations. Financial markets are inherently non-stationary, and the reported performance should therefore be interpreted within the evaluated settings rather than as evidence of universal generalisability. Further validation across additional asset classes, longer time periods, and different market regimes would provide a more comprehensive assessment of the framework’s robustness.

Second, the current framework provides limited insight into the mechanisms underlying individual predictions and trading recommendations. Future work could incorporate explainable AI and uncertainty estimation techniques to improve the transparency of the decision-support process. Adaptive monitoring mechanisms could also be introduced to detect distribution shifts and model degradation, enabling retraining or human review when market conditions change substantially.

Future research may further extend the framework by incorporating broader sources of financial information, including economic indicators and textual financial news. The integration of structured market data with multimodal information and emerging foundation models may enable more comprehensive financial decision support. These extensions, together with evaluation under extreme market events and practical data-quality constraints, represent promising directions for further investigation.

\section{Conclusion}

We have presented an end-to-end machine learning framework for structured investment decision support. The work is motivated by the view that financial machine learning should extend beyond standalone price prediction to consider how predictive outputs are translated into execution-aware and risk-controlled trading decisions.

Its architecture integrates multi-asset market data, large-scale feature engineering, execution-aware labelling, feature refinement, specialist predictive models, ensemble aggregation, structured report generation, cloud deployment, and automated daily execution. Rather than producing directional forecasts alone, the system generates actionable information comprising trade direction, entry price, stop-loss, take-profit, and risk-reward ratio.

Experimentally, the execution-aware formulation outperforms directional labelling, while the integrated pipeline improves on a simpler single-model baseline. The combined configuration using multi-asset representation and month-specialised temporal validation also improves forward-testing performance, although the individual contributions of these two factors were not isolated. Operational feasibility is demonstrated through cloud orchestration, automated report generation, daily report delivery, and reuse across selected assets and configurations. These findings are limited to the evaluated data, assets, configurations, and market conditions and should not be interpreted as evidence of universal performance.

Overall, the findings show that financial machine learning becomes more practically useful when prediction, execution logic, evaluation, and operational delivery are designed as a unified system. By providing a reproducible execution-aware decision-support pipeline, this work offers a step towards more practical, interpretable, and operationally deployable financial machine learning systems.

\bibliography{refs}

\end{document}